\documentclass[12pt,a4paper]{article}
\usepackage{fullpage}
\usepackage[centertags]{amsmath}
\allowdisplaybreaks[4]
\usepackage{amssymb}
\usepackage{bm}
\usepackage[dvips]{graphicx}
\usepackage{url}
\usepackage[dvips,hyperindex]{hyperref}
\newcommand{\bos}[1]{\overset{\scriptscriptstyle(-)}{#1}}
\newcommand{\boss}[2]{\ensuremath{\rlap{\kern-2.5pt\ensuremath{\overset{\scriptscriptstyle(-)}{\phantom{#1}}}}{\ensuremath{{#1}_{#2}}}}}
\begin{document}

\begin{flushright}
\begin{tabular}{l}
\texttt{arXiv:0711.4222v3}
\\
\textsf{17 September 2008}
\end{tabular}
\end{flushright}
\vspace{1cm}
\begin{center}
\Large\bfseries
Limits on $\bm{\nu_{e}}$ and $\bm{\bar\nu_{e}}$ disappearance from Gallium and reactor experiments
\\[0.5cm]
\large\normalfont
Mario A. Acero\ensuremath{^{(a,b,c)}}, Carlo Giunti\ensuremath{^{(b)}}, Marco Laveder\ensuremath{^{(d)}}
\\[0.5cm]
\normalsize\itshape
\setlength{\tabcolsep}{1pt}
\begin{tabular}{cl}
\ensuremath{(a)}
&
Dipartimento di Fisica Teorica,
Universit\`a di Torino,
\\
&
Via P. Giuria 1, I--10125 Torino, Italy
\\[0.3cm]
\ensuremath{(b)}
&
INFN, Sezione di Torino,
Via P. Giuria 1, I--10125 Torino, Italy
\\[0.3cm]
\ensuremath{(c)}
&
Laboratoire d'Annecy-le-Vieux de Physique Th\'orique LAPTH,
\\
&
Universit\'e de Savoie, CNRS/IN2P3, 74941 Annecy-le-vieux, France
\\[0.3cm]
\ensuremath{(d)}
&
Dipartimento di Fisica ``G. Galilei'', Universit\`a di Padova,
and
\\
&
INFN, Sezione di Padova,
Via F. Marzolo 8, I--35131 Padova, Italy
\end{tabular}
\end{center}
\begin{abstract}
The deficit
observed in the Gallium radioactive source experiments
is interpreted as a possible indication of
the disappearance of electron neutrinos.
In the effective framework of two-neutrino mixing we obtain
$\sin^{2}2\vartheta \gtrsim 0.03$
and
$\Delta{m}^{2} \gtrsim 0.1 \, \text{eV}^{2}$.
The compatibility of this result
with the data of the Bugey and Chooz reactor short-baseline antineutrino disappearance experiments is studied.
It is found that the Bugey data present a hint of neutrino oscillations
with
$0.02 \lesssim \sin^{2}2\vartheta \lesssim 0.08$
and
$\Delta{m}^{2} \approx 1.8 \, \text{eV}^{2}$,
which is compatible with the Gallium allowed region of the mixing parameters.
This hint persists in the combined analyses
of Bugey and Chooz data,
of Gallium and Bugey data,
and
of Gallium, Bugey, and Chooz data.
\end{abstract}

\newpage

\section{Introduction}
\label{001}
\nopagebreak

The observation of solar and very-long-baseline reactor neutrino oscillations due to the squared-mass difference
$ \Delta{m}^{2}_{\text{SOL}} = ( 7.59 \pm 0.21 ) \times 10^{-5} \, \text{eV}^{2} $
\protect\cite{0801.4589}
and the observation of atmospheric and long-baseline accelerator neutrino oscillations due to the squared-mass difference
$ \Delta{m}^{2}_{\text{ATM}} = 2.74 {}^{+0.44}_{-0.26} \times 10^{-3} \, \text{eV}^{2} $
\protect\cite{Adamson:2007gu}
give very robust evidence of three-neutrino mixing
(for reviews of the theory and phenomenology of neutrino mixing, see
Refs.~\protect\cite{Bilenky:1978nj,Bilenky:1987ty,hep-ph/9812360,hep-ph/0202058,hep-ph/0310238,hep-ph/0405172,hep-ph/0506083,hep-ph/0606054,Giunti-Kim-2007}).
There are, however,
some anomalies in the data of neutrino experiments which could
be interpreted as indications of exotic neutrino physics beyond three-neutrino mixing:
the LSND anomaly \protect\cite{hep-ex/0104049},
the Gallium radioactive source experiments anomaly \protect\cite{nucl-ex/0512041},
and the MiniBooNE low-energy anomaly
\protect\cite{0704.1500}.
In this paper we consider the anomaly observed
in the Gallium radioactive source experiments
\protect\cite{Anselmann:1995ar,Hampel:1998fc-Cr-51,Abdurashitov:1996dp,hep-ph/9803418,nucl-ex/0512041},
in which the Gallium solar neutrino detectors
GALLEX \protect\cite{Hampel:1998xg} and SAGE \protect\cite{nucl-ex/0509031}
were tested by measuring the electron neutrino flux
produced by intense artificial radioactive sources
placed inside the detectors.
The Gallium radioactive source experiments
measured a number of events smaller than expected.
This deficit
can be interpreted\footnote{
Another possible explanation is that
the theoretical
cross section of the Gallium detection process
has been overestimated \protect\cite{nucl-ex/0512041,hep-ph/0605186}.
}
as an indication of the disappearance of electron neutrinos
due to neutrino oscillations
\protect\cite{Laveder:2007zz,hep-ph/0610352,0707.4593}.
Under this hypothesis,
we analyze the data of the Gallium radioactive source experiments in the
effective framework of two-neutrino mixing,
which describes neutrino oscillations due to a $ \Delta{m}^{2} $
that is much larger than the solar and atmospheric ones
(see Refs.~\protect\cite{hep-ph/9812360,hep-ph/0202058,Giunti-Kim-2007}).
We also study the compatibility of this interpretation of
the Gallium radioactive source experiments anomaly
with the data of the Bugey \protect\cite{Declais:1995su} and Chooz \protect\cite{hep-ex/0301017}
reactor short-baseline antineutrino disappearance experiments.

\section{Gallium}
\label{002}
\nopagebreak

The GALLEX \protect\cite{Hampel:1998xg} and SAGE \protect\cite{nucl-ex/0509031}
solar neutrino detectors
(see
Refs.~\protect\cite{Bilenky:1978nj,Bilenky:1987ty,hep-ph/9812360,hep-ph/0202058,hep-ph/0310238,hep-ph/0405172,hep-ph/0506083,hep-ph/0606054,Giunti-Kim-2007})
have been tested
in so-called
"Gallium radioactive source experiments"
which consist in the detection of electron neutrinos
produced by intense artificial ${}^{51}\text{Cr}$ and ${}^{37}\text{Ar}$ radioactive sources
placed inside the detectors.

The radioactive nuclei
${}^{51}\text{Cr}$ and ${}^{37}\text{Ar}$
decay through electron capture
($ e^{-} + {}^{51}\text{Cr} \to {}^{51}\text{V} + \nu_{e} $
and
$ e^{-} + {}^{37}\text{Ar} \to {}^{37}\text{Cl} + \nu_{e} $)
emitting $\nu_{e}$ lines with the energies and branching ratios listed in Tab.~\ref{004}.
These neutrinos
were detected through the same reaction used for the detection of solar neutrinos
\protect\cite{Kuzmin-Ga-65}:
\begin{equation}
\nu_{e} + {}^{71}\text{Ga} \to {}^{71}\text{Ge} + e^{-}
\,,
\label{003}
\end{equation}
which has the low neutrino energy threshold
$ E_{\nu}^{\text{th}}({}^{71}\text{Ga}) = 0.233 \, \text{MeV} $.
The cross sections of the $\nu_{e}$ lines emitted in
${}^{51}\text{Cr}$ and ${}^{37}\text{Ar}$
decay interpolated from Tab.~II of Ref.~\protect\cite{Bahcall-Ga-97}
are listed in Tab.~\ref{004}.

\begin{table}[t!]
\begin{center}
\begin{tabular}{l|cccc|cc}
&
\multicolumn{4}{c|}{${}^{51}\text{Cr}$}
&
\multicolumn{2}{c}{${}^{37}\text{Ar}$}
\\
\hline
$E_{\nu}\,[\text{keV}]$             & $  747 $ & $  752 $ & $  427 $ & $  432 $ & $  811$ & $  813$ \\
B.R.                                & $0.8163$ & $0.0849$ & $0.0895$ & $0.0093$ & $0.902$ & $0.098$ \\
$\sigma\,[10^{-46}\,\text{cm}^{2}]$ & $ 60.8 $ & $ 61.5 $ & $ 26.7 $ & $ 27.1 $ & $ 70.1$ & $70.3 $ \\
\hline
\end{tabular}
\caption{ \label{004}
Energies ($E_{\nu}$), branching ratios (B.R.) and Gallium cross sections ($\sigma$)
of the $\nu_{e}$ lines emitted in
${}^{51}\text{Cr}$ and ${}^{37}\text{Ar}$
decay through electron capture.
The cross sections are interpolated from Tab.~II of Ref.~\protect\cite{Bahcall-Ga-97}.
}
\end{center}
\end{table}

The ratios $R$ of measured and predicted ${}^{71}\text{Ge}$
production rates in the two GALLEX ${}^{51}\text{Cr}$ radioactive source experiments\footnote{
As explained in Ref.~\protect\cite{nucl-ex/0512041},
the values of $R$ in Tab.~\ref{006} for the two GALLEX ${}^{51}\text{Cr}$ radioactive source experiments
are different from those published in Refs.~\protect\cite{Anselmann:1995ar,Hampel:1998fc-Cr-51},
because of an improved reanalysis of the data.
Similar results have been published recently in a PhD thesis \cite{Kaether:2007zz}
and discussed at the Neutrino 2008 Conference \cite{Hahn-Nu2008}:
$ R(\text{Cr1}) = 0.997 \pm 0.11 $
and
$ R(\text{Cr2}) = 0.807 {}^{+0.11}_{-0.10} $
in a standard rise-time analysis;
$ R(\text{Cr1}) = 0.953 \pm 0.11 $
and
$ R(\text{Cr2}) = 0.812 {}^{+0.10}_{-0.11} $
in a pulse-shape analysis.
We have verified that our results are stable against such small changes of the data.
},
Cr1 \protect\cite{Anselmann:1995ar} and Cr2 \protect\cite{Hampel:1998fc-Cr-51},
and
the SAGE
${}^{51}\text{Cr}$ \protect\cite{Abdurashitov:1996dp,hep-ph/9803418} and ${}^{37}\text{Ar}$ \protect\cite{nucl-ex/0512041} radioactive source experiments,
as reported in Ref.~\protect\cite{nucl-ex/0512041},
are listed in Tab.~\ref{006}.
Since the weighted average,
\protect\cite{nucl-ex/0512041}
\begin{equation}
R_{\text{Ga}}
=
0.88 \pm 0.05
\,,
\label{005}
\end{equation}
is smaller than unity by more than $2\sigma$,
it can be interpreted as an indication of the disappearance of electron neutrinos
due to neutrino oscillations
\protect\cite{Laveder:2007zz,hep-ph/0610352,0707.4593}.
The $\chi^2$ in the absence of oscillation is $8.19$ for 4 degrees of freedom,
corresponding to a 8.5\% goodness-of-fit\footnote{
The goodness-of-fit is the probability to obtain a worse fit under the assumption
that the model under consideration is correct (see Ref.~\cite{PDG-2006}).
It is the standard statistic used for the estimation of the quality of a fit
obtained with the least-squares method,
assuming the validity of the approximation in which
$\chi^{2}_{\text{min}}$ has a $\chi^2$ distribution with
$ \text{NDF} = N_{\text{D}} - N_{\text{P}} $ degrees of freedom,
where
$N_{\text{D}}$ is the number of data points and $N_{\text{P}}$ is the number of fitted parameters.
The fit is usually considered to be acceptable if the goodness-of-fit is larger than about 1\%.
}, as shown in Tab.~\ref{010}.
Therefore,
a fluctuation of the data in the case of no oscillations cannot be excluded.
However,
since from a physical point of view it is interesting to explore possible indications
of non-standard physics,
in the following we consider the case of neutrino oscillations.

\begin{table}[t!]
\begin{center}
\begin{tabular}{l|cc|cc}
&
\multicolumn{2}{c|}{GALLEX}
&
\multicolumn{2}{c}{SAGE}
\\
\hline
&
Cr1
&
Cr2
&
${}^{51}\text{Cr}$
&
${}^{37}\text{Ar}$
\\
\hline
$R$ & $ 1.00 \pm 0.10 $ & $ 0.81 \pm 0.10 $ & $ 0.95 \pm 0.12 $ & $ 0.79 \pm 0.10 $ \\
radius [m] & \multicolumn{2}{c|}{$1.9$} & \multicolumn{2}{c}{$0.7$} \\
height [m] & \multicolumn{2}{c|}{$5.0$} & \multicolumn{2}{c}{$1.47$} \\
source height [m] & $2.7$ & $2.38$ & \multicolumn{2}{c}{$0.72$} \\
\hline
\end{tabular}
\caption{ \label{006}
Ratios $R$ of measured and predicted ${}^{71}\text{Ge}$
production rates in the two GALLEX ${}^{51}\text{Cr}$ radioactive source experiments,
Cr1 \protect\cite{Anselmann:1995ar} and Cr2 \protect\cite{Hampel:1998fc-Cr-51},
and
the SAGE
${}^{51}\text{Cr}$ \protect\cite{Abdurashitov:1996dp,hep-ph/9803418} and ${}^{37}\text{Ar}$ \protect\cite{nucl-ex/0512041} radioactive source experiments,
as reported in Ref.~\protect\cite{nucl-ex/0512041}.
We give also the radii and heights of the GALLEX and SAGE cylindrical detectors
and the heights from the base of the detectors at which the radioactive sources were placed along the axes of the detectors.
}
\end{center}
\end{table}

In the effective framework of two-neutrino oscillations,
which is appropriate in the case of short-baseline oscillations generated by a squared-mass difference
much larger than
$ \Delta{m}^{2}_{\text{SOL}} $
and
$ \Delta{m}^{2}_{\text{ATM}} $
(see
Refs.~\protect\cite{hep-ph/9812360,Giunti-Kim-2007}),
the survival probability of electron neutrinos and antineutrinos
with energy $E_{\nu}$ at a distance $L$ from the source
is given by\footnote{
The symmetry under CPT transformations,
which is a characteristic of all relativistic local quantum field theories,
implies that the survival probabilities of neutrinos and antineutrinos are equal
(see Ref.~\protect\cite{Giunti-Kim-2007}).
}
\begin{equation}
P_{\boss{\nu}{e}\to\boss{\nu}{e}}(L,E_{\nu})
=
1 - \sin^{2}2\vartheta \, \sin^{2}\left( \frac{ \Delta{m}^{2} L }{ 4 E_{\nu} } \right)
\,,
\label{007}
\end{equation}
where $\vartheta$ is the mixing angle and $\Delta{m}^{2}$ is the squared-mass difference.
The fit of the data gives information on the values of the mixing parameters $\sin^{2}2\vartheta$ and $\Delta{m}^{2}$.

In our calculation, the theoretical value of
the ratio $R$ of the predicted ${}^{71}\text{Ge}$
production rates in each of the Gallium radioactive source experiments
in the cases of presence and absence of neutrino oscillations
is given by
\begin{equation}
R
=
\frac
{ \int \text{d}V \, L^{-2} \sum_{i} (\text{B.R.})_{i} \, \sigma_{i} \, P_{\nu_{e}\to\nu_{e}}(L,E_{\nu,i}) }
{ \sum_{i} (\text{B.R.})_{i} \, \sigma_{i} \int \text{d}V \, L^{-2} }
\,,
\label{008}
\end{equation}
where $i$ is the index of the $\nu_{e}$ lines emitted in
${}^{51}\text{Cr}$ or ${}^{37}\text{Ar}$,
which are listed in Tab.~\ref{004}.
The measured ratios are listed in Tab.~\ref{006},
together with the dimensions of the detectors,
which we approximate as cylindrical,
and the height from the base of each detector at which the radioactive sources were placed along the axis of the respective detector.
We averaged the neutrino path length $L$ with a Monte Carlo integration over the volume $V$ of each cylindrical detector.

\begin{figure}[t!]
\begin{center}
\setlength{\tabcolsep}{1pt}
\begin{tabular}{cc}
\includegraphics*[bb=25 147 572 702, width=0.49\textwidth]{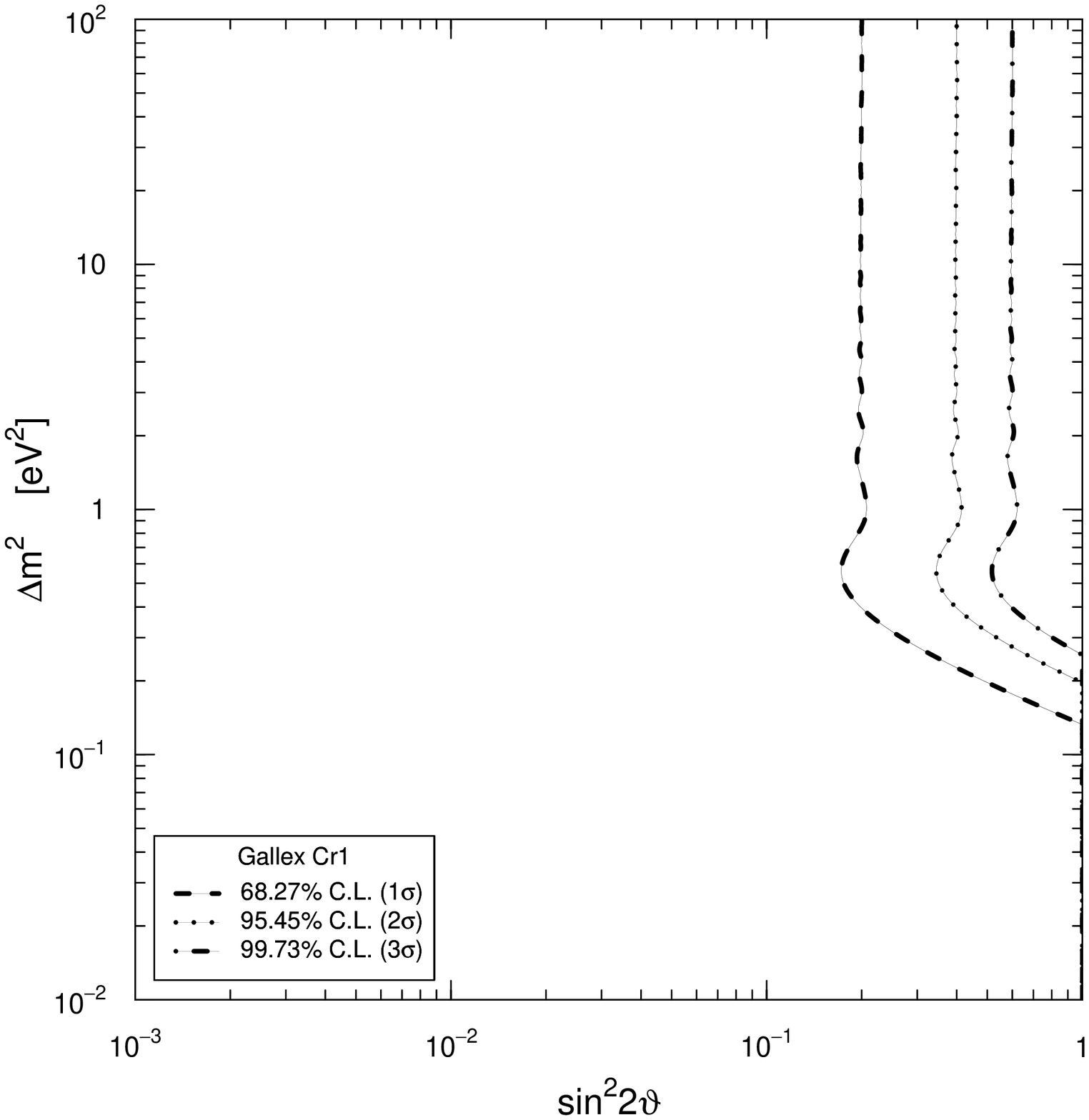}
&
\includegraphics*[bb=25 147 572 702, width=0.49\textwidth]{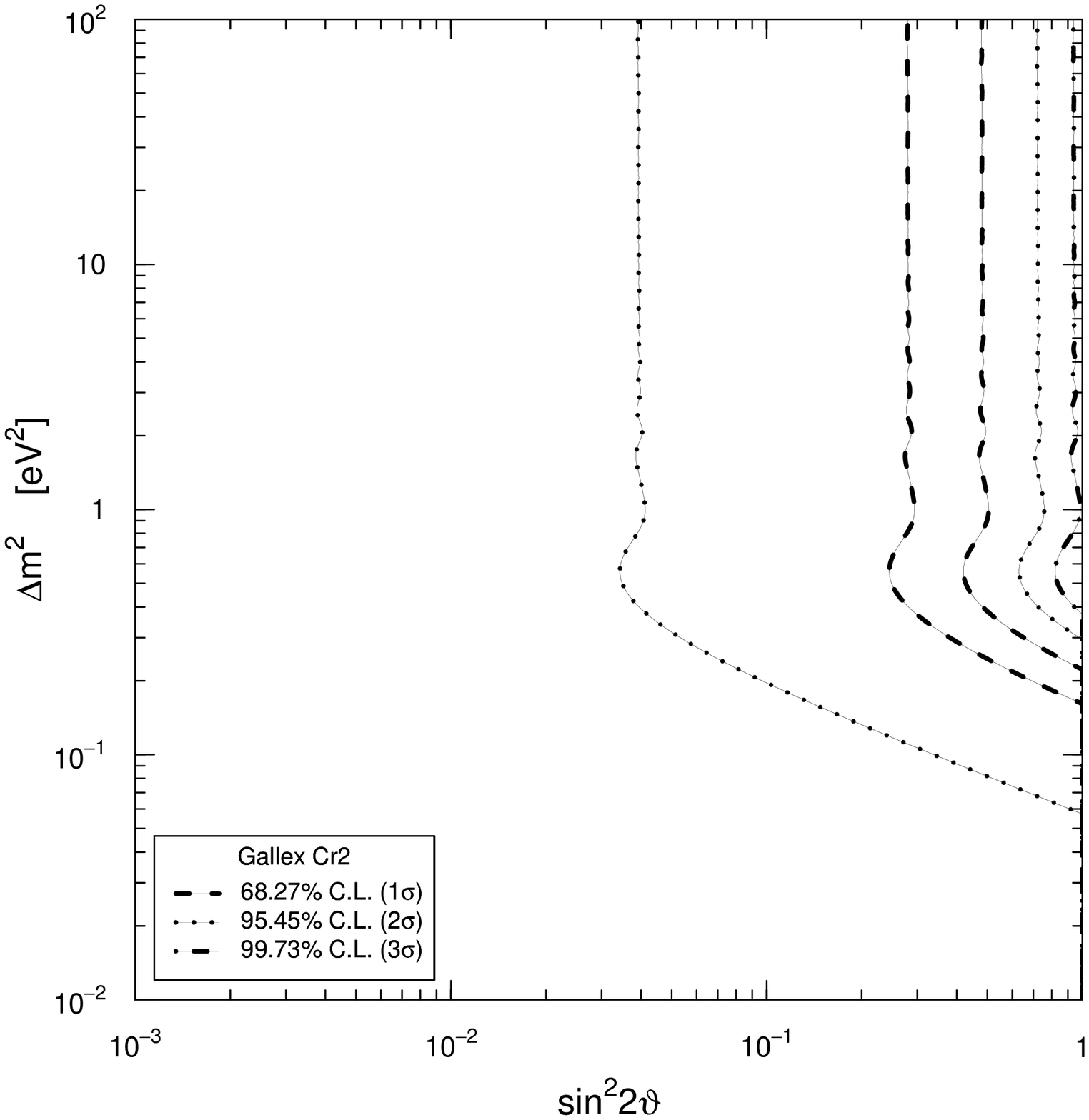}
\\
\includegraphics*[bb=25 147 572 702, width=0.49\textwidth]{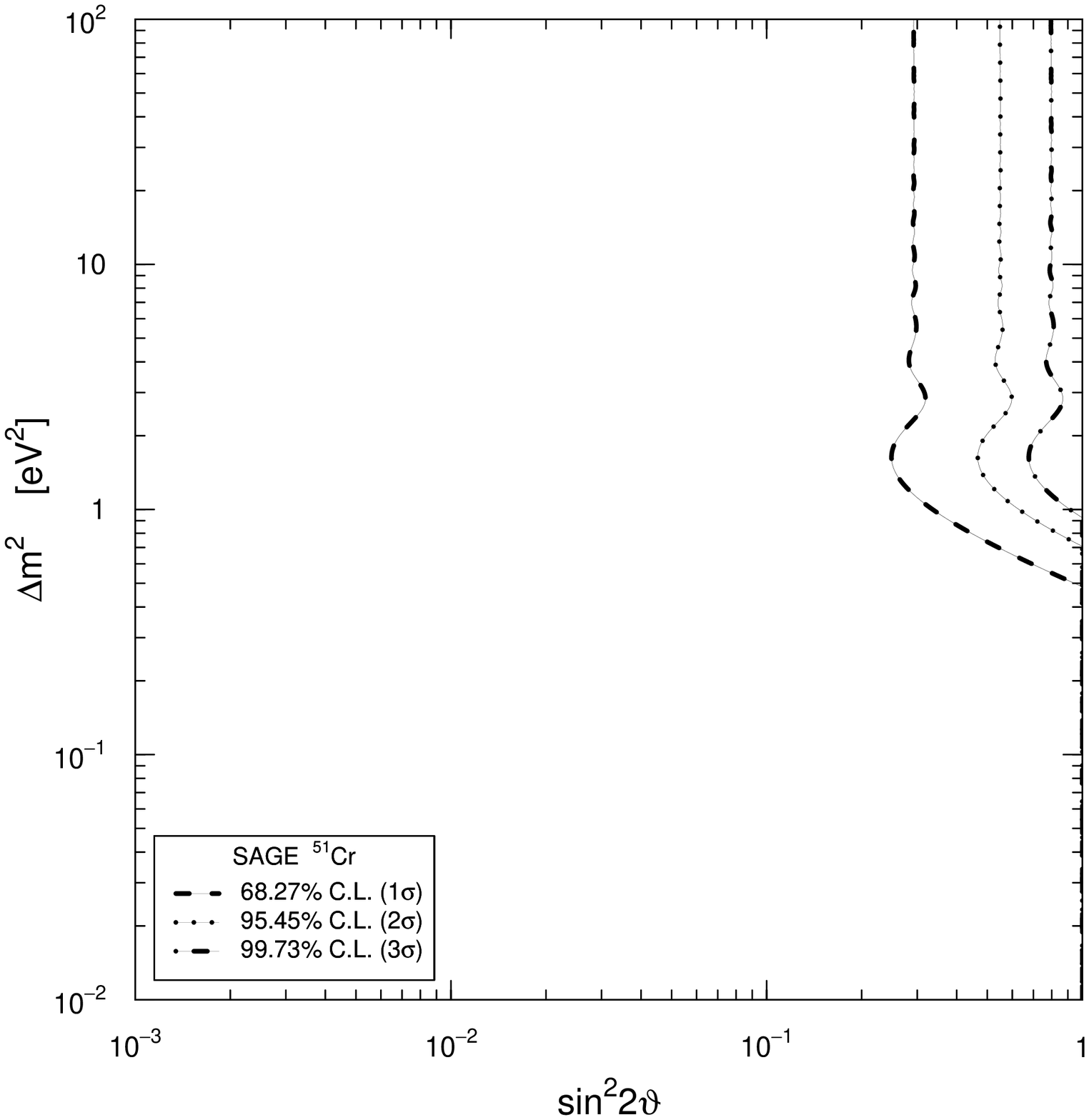}
&
\includegraphics*[bb=25 147 572 702, width=0.49\textwidth]{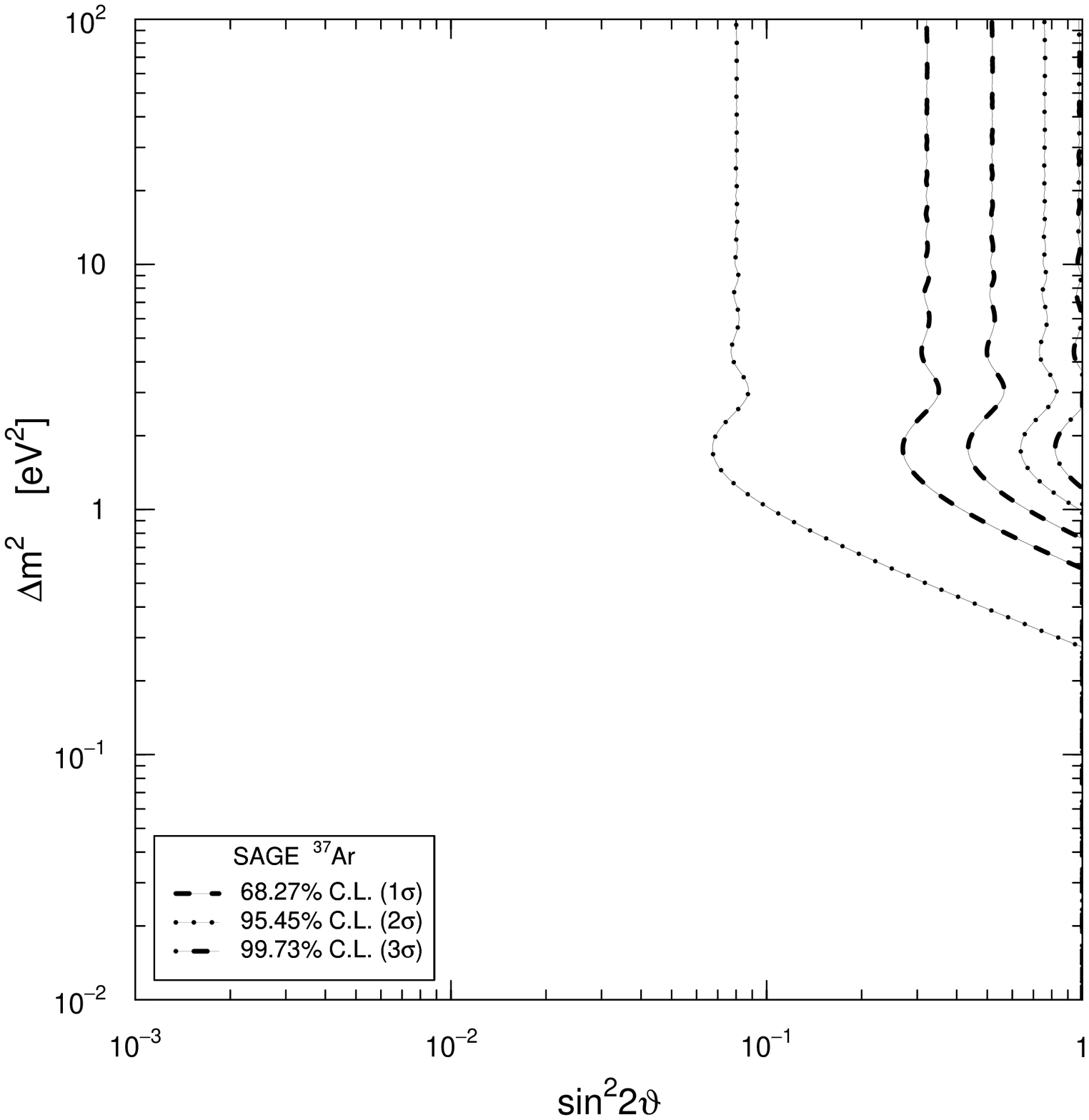}
\end{tabular}
\caption{ \label{023}
Allowed regions in the
$\sin^{2}2\vartheta$--$\Delta{m}^{2}$ plane obtained
from the fits of the results of
the two GALLEX ${}^{51}\text{Cr}$ radioactive source experiments,
Cr1 and Cr2,
and
the SAGE
${}^{51}\text{Cr}$ and ${}^{37}\text{Ar}$ radioactive source experiments.
The curves in the GALLEX Cr1 and SAGE ${}^{51}\text{Cr}$ plots exclude the region on the right.
In the GALLEX Cr2 and SAGE ${}^{37}\text{Ar}$ plots,
the pairs of $1\sigma$ and $2\sigma$ curves delimit allowed regions,
whereas the $3\sigma$ curves exclude the region on the right.
}
\end{center}
\end{figure}

\begin{table}[t!]
\begin{center}
\begin{tabular}{cccccccc}
&
&
Ga
&
Bu
&
Ga+Bu
&
Bu+Ch
&
Ga+Ch
&
Ga+Bu+Ch
\\
\hline
	&	$\chi^{2}_{\text{min}}$				&	$	8.19	$	&	$	50.94	$	&	$	59.13	$	&	$	51.00	$	&	$	8.26	$	&	$	59.19	$	\\
No Osc.	&	NDF						&	$	4	$	&	$	55	$	&	$	59	$	&	$	56	$	&	$	5	$	&	$	60	$	\\
	&	GoF						&	$	0.085	$	&	$	0.63	$	&	$	0.47	$	&	$	0.66	$	&	$	0.14	$	&	$	0.51	$	\\
\hline	&	$\chi^{2}_{\text{min}}$				&	$	2.91	$	&	$	47.97	$	&	$	53.87	$	&	$	48.63	$	&	$	6.60	$	&	$	54.80	$	\\
	&	NDF						&	$	2	$	&	$	53	$	&	$	57	$	&	$	54	$	&	$	3	$	&	$	58	$	\\
Osc.	&	GoF						&	$	0.23	$	&	$	0.67	$	&	$	0.59	$	&	$	0.68	$	&	$	0.086	$	&	$	0.60	$	\\
	&	$\sin^{2}2\vartheta_{\text{bf}} $		&	$	0.22	$	&	$	0.048	$	&	$	0.062	$	&	$	0.041	$	&	$	0.08	$	&	$	0.054	$	\\
	&	$\Delta{m}^{2}_{\text{bf}}\,[\text{eV}^{2}]$	&	$	1.98	$	&	$	1.85	$	&	$	1.85	$	&	$	1.85	$	&	$	1.72	$	&	$	1.85	$	\\
\hline	&	$\Delta\chi^{2}_{\text{min}}$				&&&	$	2.98	$	&	$	0.59	$	&	$	3.63	$	&	$	3.85	$	\\
PG	&	NDF						&&&	$	2	$	&	$	1	$	&	$	1	$	&	$	3	$	\\
	&	GoF						&&&	$	0.23	$	&	$	0.44	$	&	$	0.057	$	&	$	0.28	$	\\
\hline
\end{tabular}
\caption{ \label{010}
Values of
$\chi^{2}_{\text{min}}$,
number of degrees of freedom (NDF) and
goodness-of-fit (GoF)
for the fit of different combinations of
the results of the Gallium radioactive source experiments and the
Bugey and Chooz reactor experiments.
The first three lines correspond to the case of no oscillations (No Osc.).
The following five lines,
including the best-fit values of
$\sin^{2}2\vartheta$ and $\Delta{m}^{2}$,
correspond to the case of oscillations (Osc.).
The last three lines describe the parameter goodness-of-fit (PG) \protect\cite{hep-ph/0304176}.
}
\end{center}
\end{table}

In the separate analysis of the result of each Gallium radioactive source experiment
in terms of neutrino oscillations,
the two mixing parameters cannot be determined
through a least-squares analysis from one data point.
Therefore,
we adopt a Bayesian approach,
as done in Ref.~\protect\cite{hep-ph/9411414},
considering $R$ as a random variable with a uniform prior probability distribution
between zero and one.
Then,
if $R_{\text{obs}}$ is the observed value of $R$,
the normalized posterior probability distribution of $R$ is given by
\begin{equation}
p(R|R_{\text{obs}})
=
\frac
{ p(R_{\text{obs}}|R) }
{ \int_{0}^{1} \text{d}R \, p(R_{\text{obs}}|R) }
\,.
\label{009}
\end{equation}
Here, $p(R_{\text{obs}}|R)$ is the sampling distribution of $R_{\text{obs}}$ given $R$,
which we assume to be Gaussian
with standard deviation equal to the experimental uncertainty.
The allowed interval of $R$ with a given Bayesian Confidence Level
is given by the Highest Posterior Density interval with integrated probability equal
to the Confidence Level.
Figure~\ref{023} shows
the resulting allowed regions in the
$\sin^{2}2\vartheta$--$\Delta{m}^{2}$ plane.
One can see that the first GALLEX source experiment (Cr1)
and the ${}^{51}\text{Cr}$ SAGE source experiment,
in which the measured rate is within $1\sigma$ from unity,
imply only upper limits for the mixing parameters.
On the other hand,
the analyses of the second GALLEX source experiment (Cr2)
and the ${}^{37}\text{Ar}$ SAGE source experiment
give $2\sigma$ allowed bands,
which have a large overlap for $ \Delta{m}^{2} \gtrsim 1 \, \text{eV}^{2} $.

Let us now discuss the combined fit of the four Gallium source experiments.
Since there are enough data points to determine the two mixing parameters
$\sin^{2}2\vartheta$ and $\Delta{m}^{2}$,
we abandon the Bayesian approach in favor of a standard
frequentist least-squares fit.
This method is
based on a global minimization of the $\chi^{2}$ in the
$\sin^{2}2\vartheta$--$\Delta{m}^{2}$ plane and the calculation
of the Confidence Level contours corresponding to a
$\Delta\chi^{2}$ with two degrees of freedom:
$\Delta\chi^{2}=2.30,6.18,11.83$ for
68.27\% ($1\sigma$), 95.45\% ($2\sigma$) and 99.73\% ($3\sigma$) C.L., respectively
(see Ref.~\cite{PDG-2006}).

\begin{figure}[t!]
\begin{center}
\includegraphics*[bb=23 144 572 704, width=\textwidth]{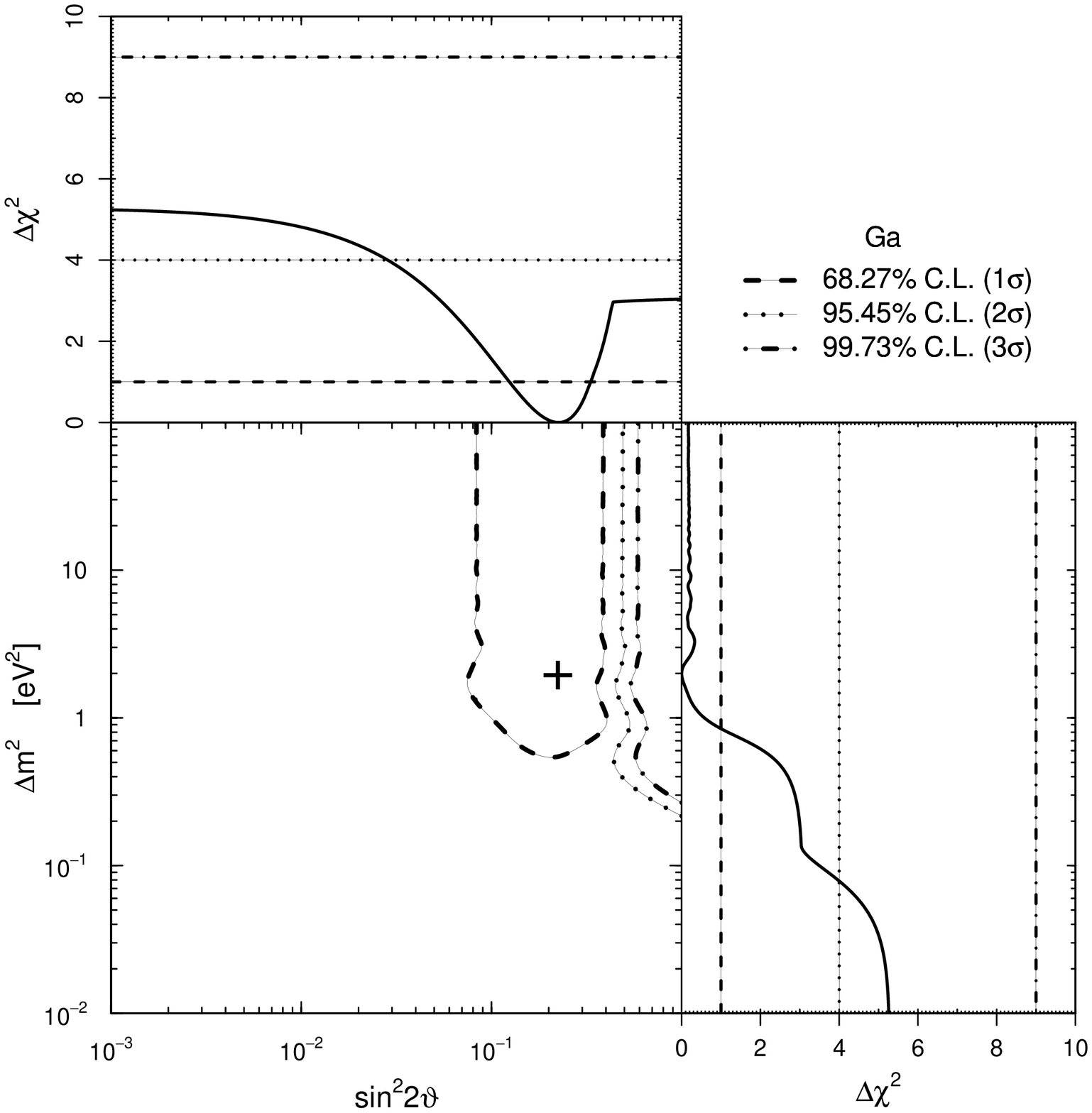}
\caption{ \label{024}
Allowed regions in the
$\sin^{2}2\vartheta$--$\Delta{m}^{2}$ plane
and
marginal $\Delta\chi^{2}$'s
for
$\sin^{2}2\vartheta$ and $\Delta{m}^{2}$
obtained from the
combined fit of the results of
the two GALLEX ${}^{51}\text{Cr}$ radioactive source experiments
and
the SAGE
${}^{51}\text{Cr}$ and ${}^{37}\text{Ar}$ radioactive source experiments.
The best-fit point corresponding to $\chi^2_{\text{min}}$ is indicated by a cross.
}
\end{center}
\end{figure}

The result of the combined least-squares analysis of the four Gallium source experiments
is shown in Fig.~\ref{024}.
One can see that there is an allowed region in the
$\sin^{2}2\vartheta$--$\Delta{m}^{2}$ plane at $1\sigma$
for
$ \Delta{m}^{2} \gtrsim 0.6 \, \text{eV}^{2} $
and
$ 0.08 \lesssim \sin^{2}2\vartheta \lesssim 0.4 $.
The values of
$\chi^{2}_{\text{min}}$,
the number of degrees of freedom (NDF),
the goodness-of-fit
(GoF)
and
the best-fit values of the mixing parameters are given in Tab.~\ref{010}.
The value of the the goodness-of-fit (23\%) shows that the fit is acceptable.

Table~\ref{011} shows the allowed ranges of
$\sin^{2}2\vartheta$ and $\Delta{m}^{2}$
obtained from the corresponding marginal $\Delta\chi^{2}\equiv\chi^{2}-\chi^{2}_{\text{min}}$ in Fig.~\ref{024}.
The presence of $2\sigma$ lower limits for $\sin^{2}2\vartheta$ and $\Delta{m}^{2}$
in spite of the absence of a $2\sigma$ lower limit in
the $\sin^{2}2\vartheta$--$\Delta{m}^{2}$ plane in Fig.~\ref{024}
is an effect due to the statistical analysis:
for one parameter $2\sigma$  corresponds to $\Delta\chi^{2}=4$,
whereas for two parameters it corresponds to $\Delta\chi^{2}=6.18$.
Hence,
it is fair to conclude that there is an indication of a possible neutrino disappearance
due to neutrino oscillations with
$\sin^{2}2\vartheta \gtrsim 0.03$
and
$\Delta{m}^{2} \gtrsim 0.1 \, \text{eV}^{2}$
at a confidence level between one and two sigmas
($ \sim 70 - 90 \% \, \text{C.L.} $).

\begin{table}[t!]
\begin{center}
\begin{tabular}{clccc}
Parameter
&
\null \hfill C.L. \hfill \null
&
Ga
&
Bu
&
Ga+Bu
\\
\hline
			& 68.27\% ($1\sigma$)	&	$		0.12	-		0.33	$	&	$		0.021	-		0.075	$	&	$		0.035	-		0.087	$	\\
$\sin^{2}2\vartheta$	& 95.45\% ($2\sigma$)	&	$	>	0.028				$	&	$			-			$	&	$		0.007	-		0.19	$	\\
			& 99.73\% ($3\sigma$)	&	$			-			$	&	$			-			$	&	$			-			$	\\
\hline
					& 68.27\% ($1\sigma$)	&	$	>	0.85				$	&	$		1.77	-		1.91	$	&	$		1.79	-		1.91	$	\\
$\Delta{m}^{2}\,[\text{eV}^{2}]$	& 95.45\% ($2\sigma$)	&	$	>	0.079				$	&	$			-			$	&	$	>	0.77				$	\\
					& 99.73\% ($3\sigma$)	&	$			-			$	&	$			-			$	&	$			-			$	\\
\hline
\end{tabular}
\caption{ \label{011}
Allowed ranges of
$\sin^{2}2\vartheta$ and $\Delta{m}^{2}$
from the combined fit of the results of Gallium radioactive source experiments,
from the fit of the results of the Bugey reactor experiment,
and
from the combined fit.
The dash indicates the absence of limits.
}
\end{center}
\end{table}

\section{Bugey}
\label{012}
\nopagebreak

The disappearance of electron antineutrinos have been investigated
by several reactor neutrino experiments at different baselines
(see Refs.~\protect\cite{hep-ph/0107277,Giunti-Kim-2007}).
Since,
according to Eq.~(\ref{007}),
the survival probabilities of neutrinos and antineutrinos are equal,
the interpretation of the results of Gallium radioactive source experiments
in terms of electron neutrino disappearance can be compared directly with the
results of reactor neutrino experiments.

In this section we consider the results of the reactor short-baseline Bugey experiment \protect\cite{Declais:1995su},
which put the most stringent constraints on the
disappearance of electron antineutrinos due to
$ \Delta{m}^{2} \gtrsim 0.1 \, \text{eV}^{2} $.

Reactor neutrino experiments detect electron antineutrinos
through the inverse neutron decay process
\begin{equation}
\bar\nu_{e} + p \to n + e^{+}
\,.
\label{013}
\end{equation}
The neutrino energy $E_{\nu}$ and the positron kinetic energy $T_{e}$ are related by
\begin{equation}
E_{\nu} = T_{e} + T_{n} + m_{e} + m_{n} - m_{p} \simeq T_{e} + 1.8 \, \text{MeV}
\,,
\label{014}
\end{equation}
where $T_{n}$ is the negligibly small recoil kinetic energy of the neutron.
In the Bugey experiment the survival probability of electron antineutrinos
was measured at three source-detector distances:
$ L_{j} = 15, 40, 95 \, \text{m} $, for $j=1,2,3$, respectively.
We use the ratio of observed and expected
(in the case of no oscillation)
positron spectra given in Fig.~17 of Ref.~\protect\cite{Declais:1995su},
in which there are
$ N_{j} = 25, 25, 10$
energy bins.
We analyze the data with the following $\chi^{2}$,
taken from Ref.~\protect\cite{Declais:1995su}:
\begin{equation}
\chi^{2}
=
\sum_{j=1}^{3}
\left\{
\sum_{i=1}^{N_{j}}
\dfrac{ \left[ \left( A a_{j} + b \left( E_{ji} - E_{0} \right) \right) R_{ji}^{\text{the}} - R_{ji}^{\text{exp}} \right]^{2} }{ \sigma_{ji}^{2} }
+
\dfrac{ \left( a_{j} - 1 \right)^{2} }{ \sigma_{a_{j}}^{2} }
\right\}
+
\dfrac{ \left( A - 1 \right)^{2} }{ \sigma_{A}^{2} }
+
\dfrac{ b^{2} }{ \sigma_{b}^{2} }
\,,
\label{015}
\end{equation}
where
$E_{ji}$ is the central energy of
the $i\text{th}$ bin in the positron kinetic energy spectrum measured at the $L_{j}$ source-detector distance,
$R_{ji}^{\text{exp}}$ and $R_{ji}^{\text{the}}$
are, respectively, the corresponding measured and calculated ratios.
The uncertainties $\sigma_{ji}$ include the statistical uncertainty of each bin
and a 1\% systematic uncertainty added in quadrature,
which takes into account the uncertainty of the spectrum calculation
(with a total of about 5\% uncorrelated systematic uncertainty over 25 bins).
The coefficients
$ \left( A a_{j} + b \left( E_{ji} - E_{0} \right) \right) $,
with $E_{0} = 1 \, \text{MeV}$,
were introduced in Ref.~\protect\cite{Declais:1995su}
in order to take into account the systematic uncertainty of the positron energy calibration.
The value of $\chi^{2}$ as a function of
$\sin^{2}2\vartheta$ and $\Delta{m}^{2}$
is calculated by minimizing Eq.~(\ref{015}) with respect to the five parameters
$A$, $a_{j}$ ($j=1,2,3$), $b$, which have, respectively,
uncertainties
$\sigma_{A} = 0.048$,
$\sigma_{a_{j}} = 0.014$
$\sigma_{b} = 0.02 \, \text{MeV}^{-1}$ \protect\cite{Declais:1995su}.
Following Ref.~\protect\cite{hep-ph/0102252},
we approximate
the neutrino flux, the detection cross section and the detection efficiency
as constants in each energy bin.
Then,
$R_{ji}^{\text{the}}$ is given by
\begin{equation}
R_{ji}^{\text{the}}
=
\frac
{
\int \text{d}L
\,
L^{-2}
\int_{E_{ji}-\Delta{E_{j}}/2}^{E_{ji}+\Delta{E_{j}}/2} \text{d}E
\int_{-\infty}^{+\infty} \text{d}T_{e}
\,
F(E,T_{e})
\,
P_{\bar\nu_{e}\to\bar\nu_{e}}(L,E_{\nu}) }
{
\Delta{E_{j}}
\int \text{d}L \, L^{-2}
}
\,.
\label{016}
\end{equation}
Here
$T_{e}$ and $E_{\nu}$ are, respectively, the positron kinetic energy and the neutrino energy,
related by Eq.~(\ref{014}),
whereas $E$ is the measured positron kinetic energy,
which is connected to $T_{e}$ by the energy resolution function of the detector $F(E,T_{e})$.
We considered a Gaussian energy resolution function with standard deviation
$0.252\sqrt{E/4.2\text{MeV}}\,\text{MeV}$
\protect\cite{Declais:1995su}.
The quantities $ \Delta{E_{j}} $ are the widths of the energy bins in each detector.
The integration over the neutrino path length $L$ is performed by a Monte Carlo
which takes into account the geometries of
the reactor and of the detectors and their relative positions \protect\cite{Declais-2008}.

\begin{figure}[t!]
\begin{center}
\includegraphics*[bb=25 147 564 702, width=\textwidth]{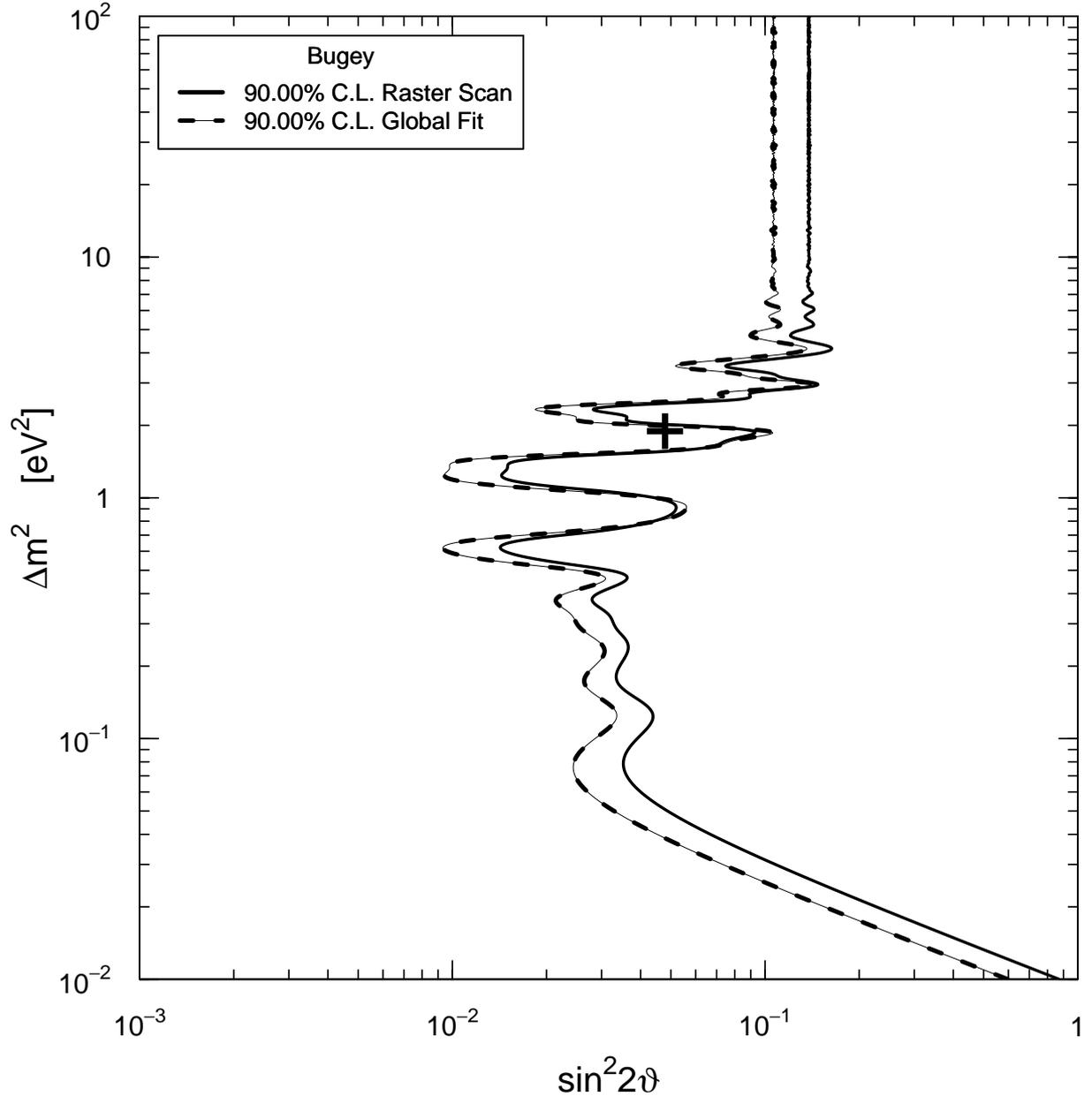}
\caption{ \label{025}
90\% C.L. exclusion curves in the
$\sin^{2}2\vartheta$--$\Delta{m}^{2}$ plane
obtained from a raster-scan analysis of Bugey data (solid line)
and from a standard global least-squares fit (dashed line).
}
\end{center}
\end{figure}

With this method we obtained the 90\% C.L. raster-scan\footnote{
In the raster-scan method,
$\chi^{2}_{\text{min}}$ is found for each fixed value of $\Delta{m}^{2}$.
The corresponding upper limit for $\sin^{2}2\vartheta$ is calculated
as the value of $\sin^{2}2\vartheta$ for which
the cumulative distribution function of
$\Delta\chi^{2}\equiv\chi^{2}-\chi^{2}_{\text{min}}$, which has one degree of freedom,
is equal to the Confidence Level
($\Delta\chi^{2}=2.71$ for 90\% C.L.).
}
exclusion curve
shown in Fig.~\ref{025},
which is similar to the original 90\% C.L. raster-scan Bugey exclusion curve in Ref.~\protect\cite{Declais:1995su}.

\begin{figure}[t!]
\begin{center}
\includegraphics*[bb=23 144 572 704, width=\textwidth]{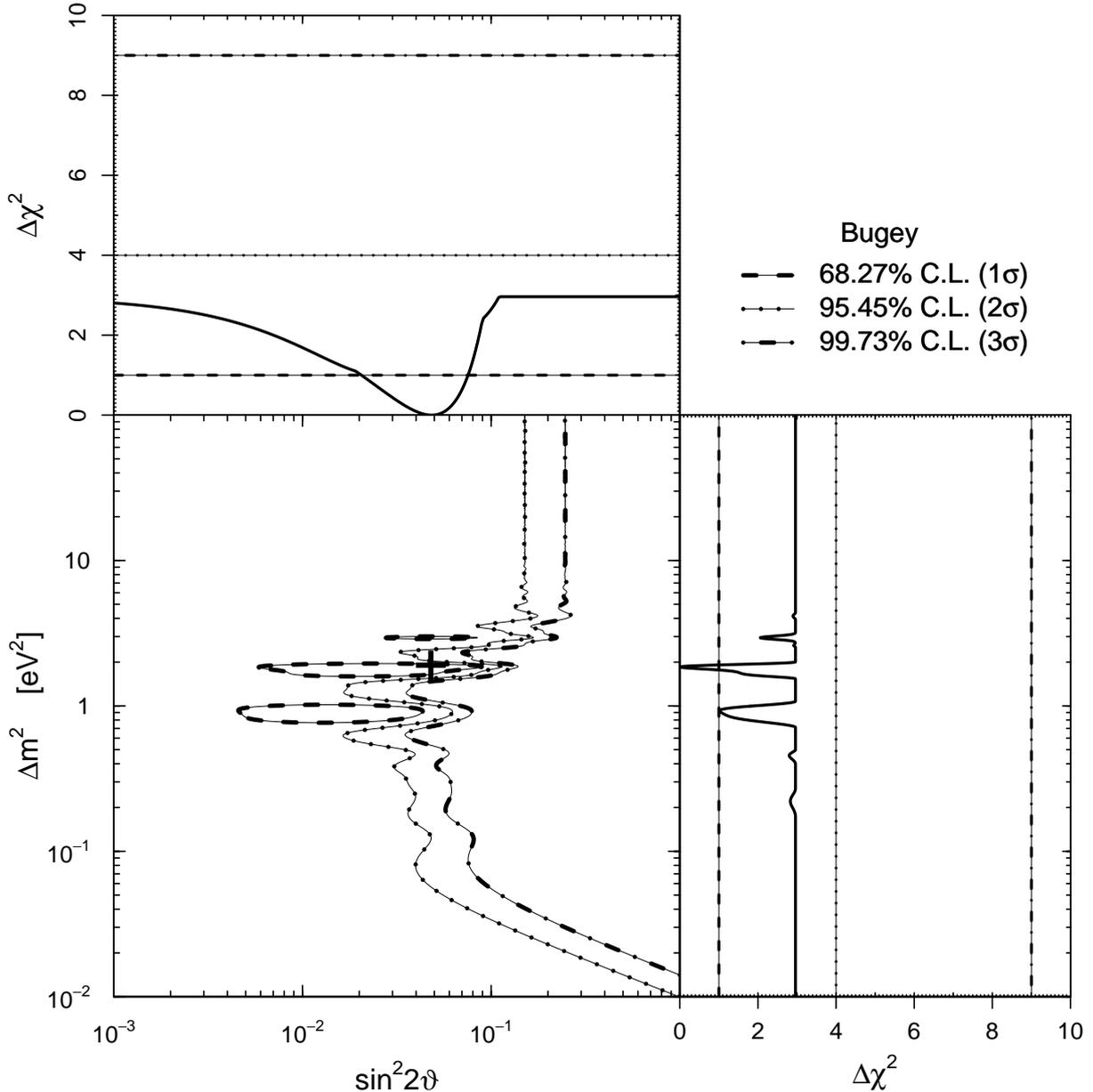}
\caption{ \label{026}
Allowed regions in the
$\sin^{2}2\vartheta$--$\Delta{m}^{2}$ plane
and
marginal $\Delta\chi^{2}$'s
for
$\sin^{2}2\vartheta$ and $\Delta{m}^{2}$
obtained from the
least-squares analysis of Bugey data.
The best-fit point corresponding to $\chi^2_{\text{min}}$ is indicated by a cross.
}
\end{center}
\end{figure}

Let us emphasize that the raster-scan method is statistically weak,
because it does not have proper coverage \protect\cite{physics/9711021}.
We presented in Fig.~\ref{025} the raster-scan exclusion curve only to show by comparison
with the analogous figure in Ref.~\protect\cite{Declais:1995su} that
our analysis of the Bugey data is acceptable.
The dashed line in Fig.~\ref{025} shows the 90\% C.L. Bugey exclusion curve
obtained with the standard least-squares method,
which we adopted also in the previous Fig.~\ref{024} and the following Figs.~\ref{026}--\ref{031}.
From Fig.~\ref{025} one can see that the 90\% C.L. raster-scan exclusion curve
overcovers for all values of $\Delta{m}^2$,
except for small intervals around
$ \Delta{m}^2 \simeq 0.9 \, \text{eV}^2 $
and
$ \Delta{m}^2 \simeq 1.9 \, \text{eV}^2 $.

\begin{figure}[t!]
\begin{center}
\includegraphics*[bb=28 142 571 700, width=0.8\textwidth]{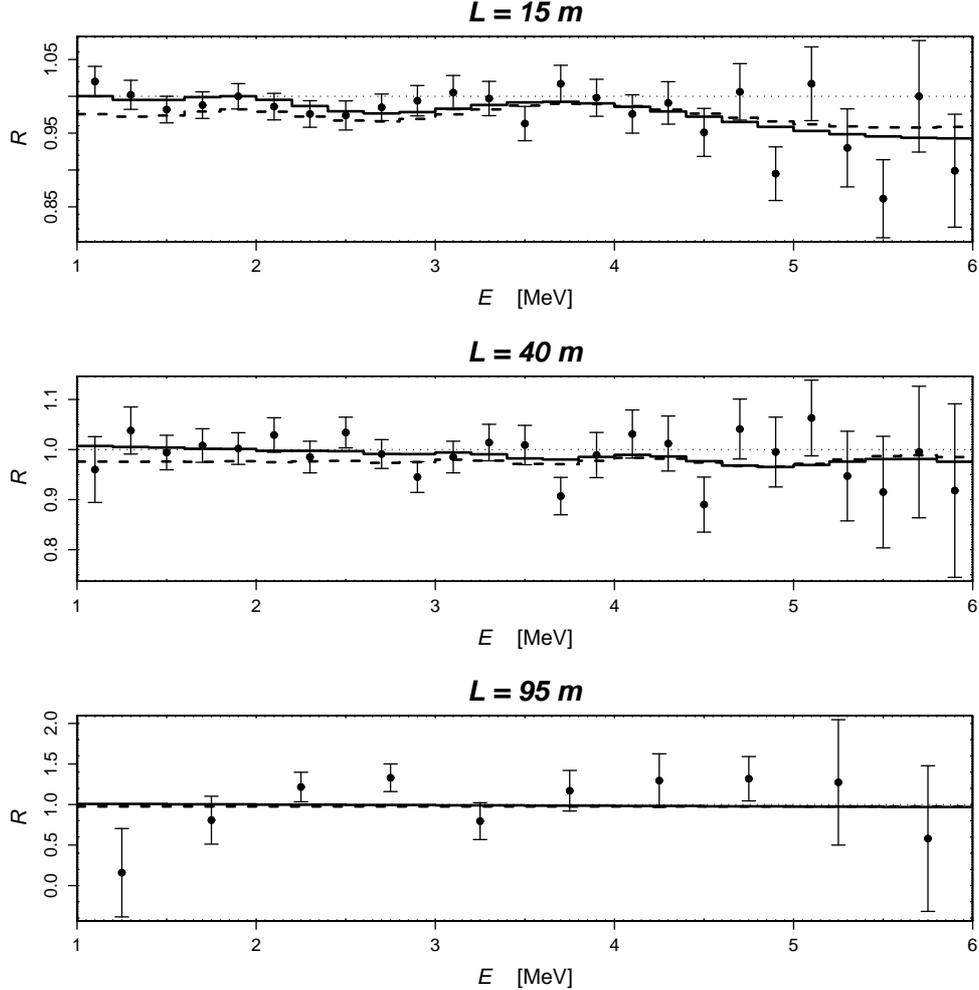}
\caption{ \label{027}
Best fit of Bugey data (points with error bars \protect\cite{Declais:1995su}).
The three panels show the ratio $R$ of observed and expected
(in the case of no oscillation) event rates
at the
three source-detector distances in the Bugey experiment
as functions of the measured positron kinetic energy $E$ (see Eq.~(\ref{016})).
In each panel,
the solid and dashed histograms correspond, respectively, to the best-fit values of
$ \left( A a_{j} + b \left( E_{ji} - E_{0} \right) \right) R_{ji}^{\text{the}} $
and
$ R_{ji}^{\text{the}} $
(see Eq.~(\ref{015})).
}
\end{center}
\end{figure}

Figure~\ref{026}
shows the allowed regions in the
$\sin^{2}2\vartheta$--$\Delta{m}^{2}$ plane
and
the marginal $\Delta\chi^{2}$'s
for
$\sin^{2}2\vartheta$ and $\Delta{m}^{2}$
obtained from the
least-squares analysis of Bugey data.
The value and location in the $\sin^{2}2\vartheta$--$\Delta{m}^{2}$ plane
of the minimum of the $\chi^{2}$,
the number of degrees of freedom (NDF) and the goodness-of-fit (GoF) are given in Tab.~\ref{010}.
The fit is satisfactory,
since the goodness-of-fit is 67\%.
The best-fit value of the oscillation parameters
and the small $1\sigma$ allowed regions in Fig.~\ref{026}
are in favor of neutrino oscillations.
However,
the $2\sigma$ and $3\sigma$ contours in Fig.~\ref{026}
provide only upper limits to neutrino oscillations.
Also,
the value of the $\chi^{2}$ in the case of absence of oscillations
and the corresponding goodness-of-fit
(63\%)
do not allow us to exclude the absence of oscillations.

The reason of the hint in favor
of neutrino oscillations given by the Bugey data
is illustrated in Fig.~\ref{027},
where
the histogram relative to the best fit is shown against
the Bugey $R_{ji}^{\text{exp}}$'s.
With the help of the histogram,
one can see that there is a weak hint of oscillations.
The $1\sigma$ allowed regions in Fig.~\ref{026} have very narrow
$\Delta{m}^{2}$ ranges around
$0.9\,\text{eV}^{2}$,
$1.85\,\text{eV}^{2}$, and
$3\,\text{eV}^{2}$,
because slight shifts of $\Delta{m}^{2}$ from these optimal values
spoil the agreement with the data of the histogram in Fig.~\ref{027}.

Table~\ref{011} shows the marginal allowed ranges of
$\sin^{2}2\vartheta$ and $\Delta{m}^{2}$
obtained from the corresponding $\Delta\chi^{2}$'s in Fig.~\ref{026}.
One can see that there is a hint of neutrino oscillations
with
$0.02 \lesssim \sin^{2}2\vartheta \lesssim 0.08$
and
$\Delta{m}^{2} \approx 1.8 \, \text{eV}^{2}$.

\begin{figure}[t!]
\begin{center}
\includegraphics*[bb=23 144 572 704, width=\textwidth]{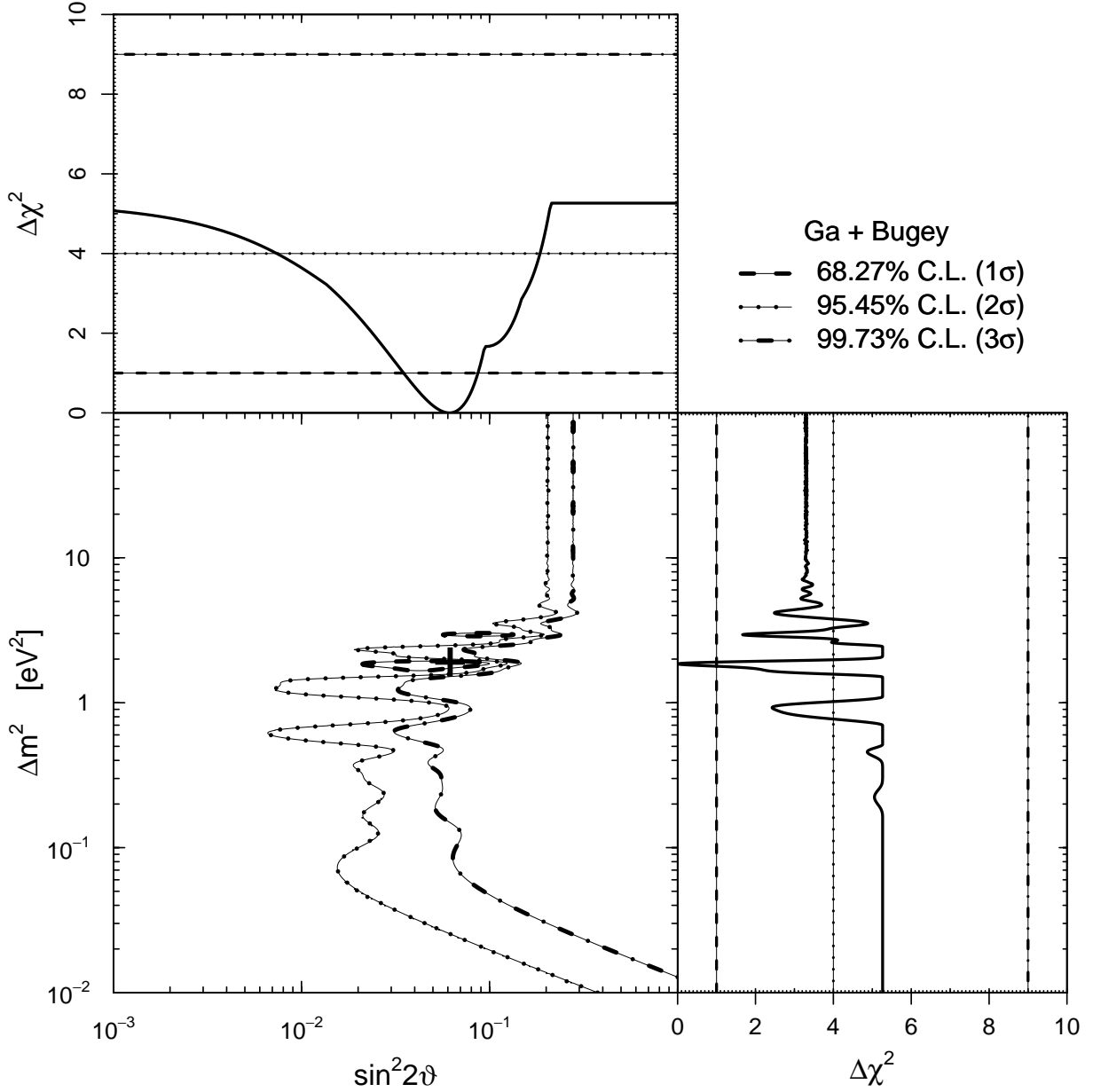}
\caption{ \label{028}
Allowed regions in the
$\sin^{2}2\vartheta$--$\Delta{m}^{2}$ plane
and
marginal $\Delta\chi^{2}$'s
for
$\sin^{2}2\vartheta$ and $\Delta{m}^{2}$
obtained from the
combined fit of the results of
the two GALLEX ${}^{51}\text{Cr}$ radioactive source experiments,
the SAGE
${}^{51}\text{Cr}$ and ${}^{37}\text{Ar}$ radioactive source experiments
and the Bugey reactor experiment.
The best-fit point corresponding to $\chi^2_{\text{min}}$ is indicated by a cross.
}
\end{center}
\end{figure}

From a comparison of Figs.~\ref{024} and \ref{026}
one can see that the allowed regions
of the Gallium radioactive source experiments and the Bugey experiment
are marginally compatible
for
$ \sin^{2} 2\vartheta \sim 0.1 $
and
$ \Delta{m}^{2} \gtrsim 1 \, \text{eV}^{2} $.
Figure~\ref{028} shows the allowed regions obtained from the combined fit.
Since the Bugey data are statistically dominant,
the curves in Fig.~\ref{028} are not very different from those in Fig.~\ref{026},
which have been obtained from the fit of the Bugey data alone.
The inclusion of the Gallium data has the effect
of eliminating the $1\sigma$ allowed region at
$ \Delta{m}^{2} \approx 0.9 \, \text{eV}^{2} $
and
of disfavoring at $1\sigma$ values of $ \sin^{2} 2\vartheta $
smaller than about $2\times10^{-2}$.
The value and location of $\chi^{2}_{\text{min}}$,
the number of degrees of freedom and the goodness-of-fit are listed in Tab.~\ref{010}.
One can see that the Gallium data do not spoil the good fit of the Bugey data.
Indeed,
the value of the parameter goodness-of-fit\footnote{
The value of $(\Delta\chi^{2}_{\text{min}})_{\text{A+B}}$
corresponding to the parameter goodness-of-fit of two experiments A and B
is given by
$ (\chi^{2}_{\text{min}})_{\text{A+B}} - [ (\chi^{2}_{\text{min}})_{\text{A}} + (\chi^{2}_{\text{min}})_{\text{B}} ] $.
It has a $\chi^2$ distribution with number of degrees of freedom
$ \text{NDF} = P_{\text{A}} + P_{\text{B}} - P_{\text{A}+\text{B}} $,
where $P_{\text{A}}$, $P_{\text{B}}$ and $P_{\text{A}+\text{B}}$ are, respectively,
the number of parameters in the fits of A, B and A+B data
\protect\cite{hep-ph/0304176}.
}
\protect\cite{hep-ph/0304176}
reported in Tab.~\ref{010} shows that the Bugey and Gallium data are compatible
under the hypothesis of neutrino oscillations.
The marginal allowed ranges of
$\sin^{2}2\vartheta$ and $\Delta{m}^{2}$
obtained from the corresponding $\Delta\chi^{2}$'s in Fig.~\ref{028}
are given in Tab.~\ref{011}.

\section{Chooz}
\label{017}
\nopagebreak

In this section we consider the result of the long-baseline reactor neutrino experiment
Chooz \protect\cite{hep-ex/0301017},
which gives limits on neutrino oscillations which are comparable with those of the Bugey experiment for
$ \Delta{m}^{2} \gtrsim 2 \, \text{eV}^{2} $.

In the Chooz experiment
the ratio of the number of observed events and that expected in the absence of neutrino oscillations
is
\begin{equation}
R_{\text{Chooz}}
=
1.01 \pm 0.04
\,.
\label{018}
\end{equation}
The value of this ratio puts a constraint on the disappearance of
electron (anti)neutrinos with energies in the MeV range
at distances smaller than about 1 km.
This corresponds to a constraint on $ \sin^{2} 2\vartheta $
for $ \Delta{m}^{2} \gtrsim 10^{-3} \, \text{eV}^{2} $.
In the range of sensitivity of the Gallium radioactive source experiments,
$ \Delta{m}^{2} \gtrsim 10^{-1} \, \text{eV}^{2} $
(see Figs.~\ref{024}),
the oscillation length of reactor antineutrinos
is much shorter than the Chooz source-detector distance.
In this case, the Chooz experiment is only sensitive to the averaged survival probability
\begin{equation}
\langle P_{\boss{\nu}{e}\to\boss{\nu}{e}} \rangle
=
1 - \frac{1}{2} \, \sin^{2}2\vartheta
\,.
\label{019}
\end{equation}
Therefore, the Chooz result in Eq.~(\ref{018})
can be combined\footnote{
In our figures we considered $\Delta{m}^{2}$
in the range $10^{-2}-10^{2}\,\text{eV}^2$.
For simplicity, we neglected the small
$\Delta{m}^{2}$ dependence of the CHOOZ exclusion curve
for $\Delta{m}^{2}\lesssim4\times10^{-2}\,\text{eV}^2$
(see Fig.~55 of Ref.~\protect\cite{hep-ex/0301017}).
}
with the results of the Gallium radioactive source experiments
simply by considering it as a measurement of $\sin^{2}2\vartheta$:
in the Bayesian approach of Eq.~(\ref{009})
\begin{equation}
\sin^{2}2\vartheta
<
0.071 ,\,
0.15 ,\,
0.23
\,,
\label{020}
\end{equation}
at
68.27\% ($1\sigma$),
95.45\% ($2\sigma$),
99.73\% ($3\sigma$) Bayesian Confidence Level, respectively.

\begin{table}[t!]
\begin{center}
\setlength{\tabcolsep}{5pt}
\begin{tabular}{clccc}
Parameter
&
\null \hfill C.L. \hfill \null
&
Bu+Ch
&
Ga+Ch
&
Ga+Bu+Ch
\\
\hline
			& 68.27\% ($1\sigma$)	&	$		0.012	-		0.067	$	&	$		0.017	-		0.14	$	&	$		0.028	-		0.078	$	\\
$\sin^{2}2\vartheta$	& 95.45\% ($2\sigma$)	&	$				<	0.096	$	&	$				<	0.20	$	&	$		0.002	-		0.12	$	\\
			& 99.73\% ($3\sigma$)	&	$				<	0.18	$	&	$				<	0.26	$	&	$				<	0.18	$	\\
\hline
					& 68.27\% ($1\sigma$)	&	$		0.83	-		1.92	$	&	$	>	0.62				$	&	$		1.78	-		1.91	$	\\
$\Delta{m}^{2}\,[\text{eV}^{2}]$	& 95.45\% ($2\sigma$)	&	$			-			$	&	$			-			$	&	$	>	0.74				$	\\
					& 99.73\% ($3\sigma$)	&	$			-			$	&	$			-			$	&	$			-			$	\\
\hline
\end{tabular}
\caption{ \label{021}
Allowed ranges of
$\sin^{2}2\vartheta$ and $\Delta{m}^{2}$
from the combined fit of the results of
the Bugey and Chooz reactor experiments,
the Gallium radioactive source and Chooz reactor experiments,
and
the Gallium radioactive source and the Bugey and Chooz reactor experiments.
The dash indicates the absence of limits.
}
\end{center}
\end{table}

\begin{figure}[t!]
\begin{center}
\includegraphics*[bb=23 144 572 704, width=\textwidth]{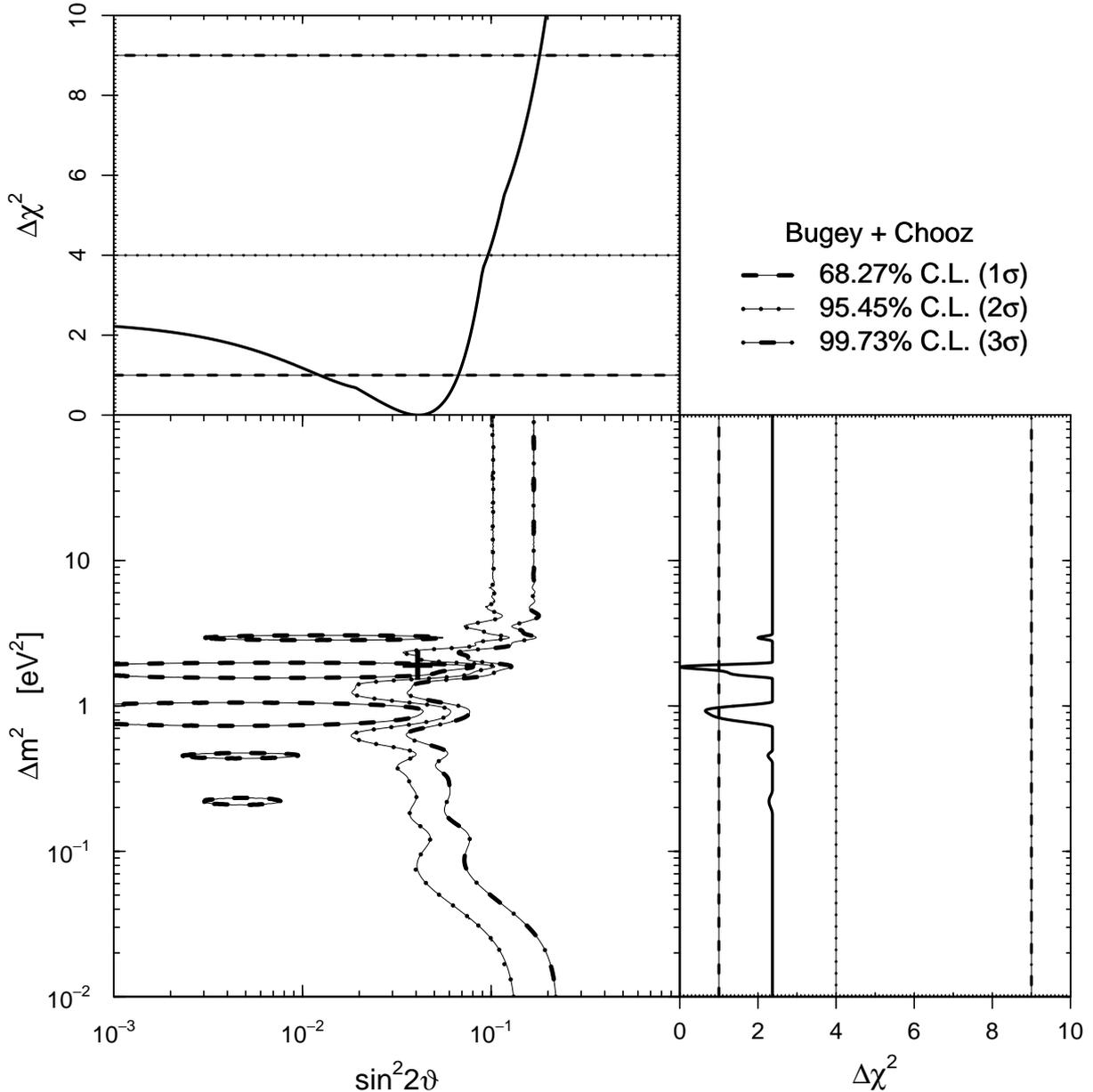}
\caption{ \label{029}
Allowed regions in the
$\sin^{2}2\vartheta$--$\Delta{m}^{2}$ plane
and
marginal $\Delta\chi^{2}$'s
for
$\sin^{2}2\vartheta$ and $\Delta{m}^{2}$
obtained from the
combined fit of the results of
the Bugey and Chooz reactor experiments.
The best-fit point corresponding to $\chi^2_{\text{min}}$ is indicated by a cross.
}
\end{center}
\end{figure}

First,
we performed a combined frequentist least-squares analysis  of the Bugey and Chooz data,
which yielded the allowed regions in the
$\sin^{2}2\vartheta$--$\Delta{m}^{2}$ plane
shown in Fig.~\ref{029},
the best fit values of the mixing parameters reported in Tab.~\ref{010},
and the marginal allowed ranges listed in Tab.~\ref{021}.
One can see that the addition of the Chooz result to the
Bugey data analysis has the effect of improving slightly
the upper limit on $\sin^{2}2\vartheta$ for $\Delta{m}^{2} \gtrsim 3 \, \text{eV}^{2}$
and that of excluding values of $\sin^{2}2\vartheta$ larger than about 0.1
for $\Delta{m}^{2} \lesssim 3 \times 10^{-2} \, \text{eV}^{2}$,
where Bugey is not sensitive.
In the intermediate range of $\Delta{m}^{2}$,
where Bugey is sensitive to the oscillations,
the addition of the Chooz result weakens the hint in favor of oscillations
given by the Bugey data: the $1\sigma$ allowed regions in Fig.~\ref{026}
are stretched towards small values of $\sin^{2}2\vartheta$ in Fig.~\ref{029}.
However,
the best-fit value of the mixing parameters remain unchanged,
because of the dominance of the Bugey data.
From Tab.~\ref{010},
one can see that
the parameter goodness-of-fit implies that Bugey and Chooz results are compatible under the hypothesis of
neutrino oscillations,
but the goodness-of-fit obtained in the case of no oscillations do not allow us to exclude this possibility.

\begin{figure}[t!]
\begin{center}
\includegraphics*[bb=23 144 572 704, width=\textwidth]{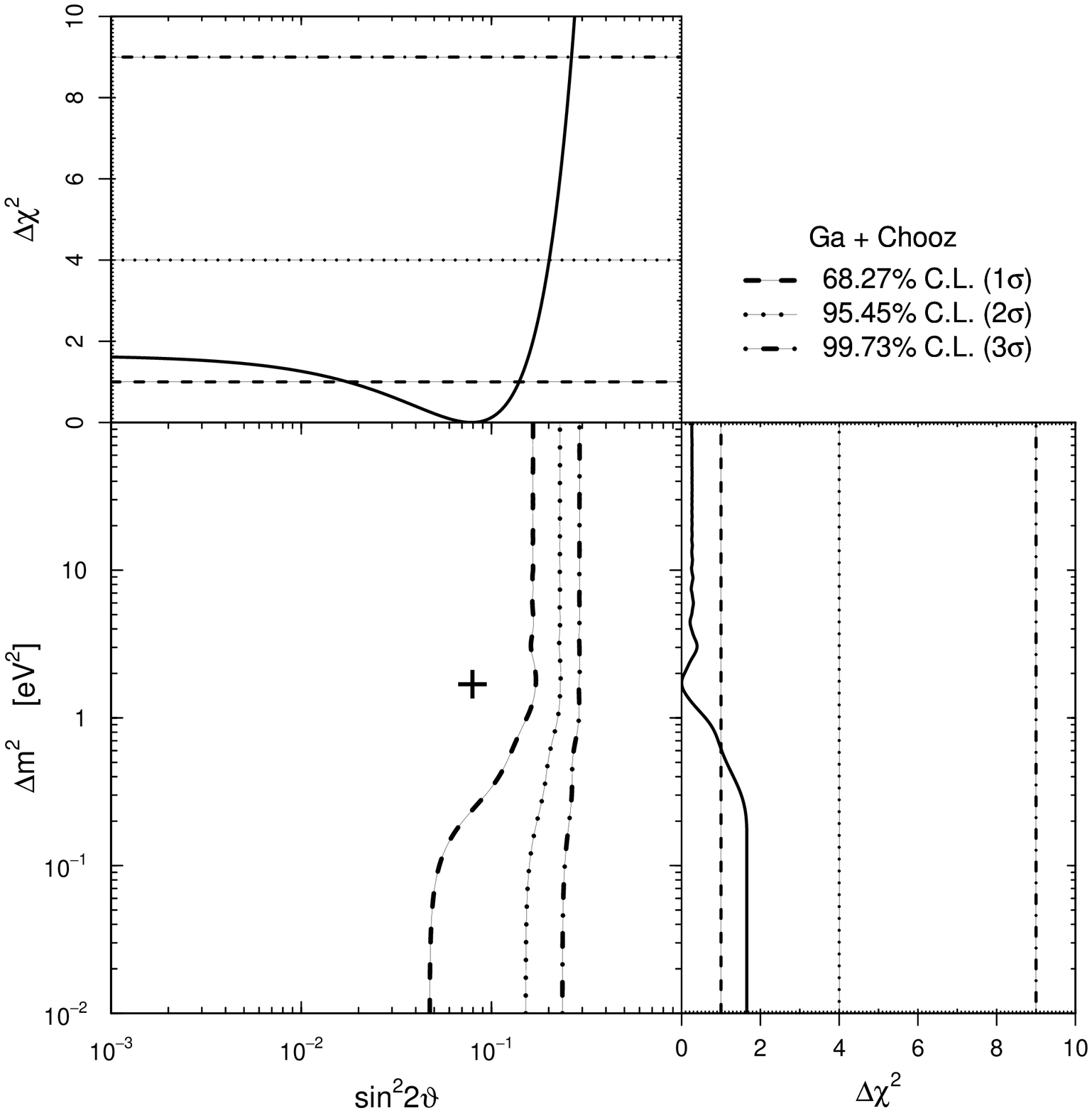}
\caption{ \label{030}
Allowed regions in the
$\sin^{2}2\vartheta$--$\Delta{m}^{2}$ plane
and
marginal $\Delta\chi^{2}$'s
for
$\sin^{2}2\vartheta$ and $\Delta{m}^{2}$
obtained from the
combined fit of the results of
the two GALLEX ${}^{51}\text{Cr}$ radioactive source experiments,
the SAGE
${}^{51}\text{Cr}$ and ${}^{37}\text{Ar}$ radioactive source experiments
and the Chooz reactor experiment.
The best-fit point corresponding to $\chi^2_{\text{min}}$ is indicated by a cross.
}
\end{center}
\end{figure}

From the comparison of Eq.~(\ref{020}) and Fig.~\ref{024},
one can see that the results of the Chooz and the Gallium radioactive source experiments
are compatible only at the $2\sigma$ level.
In fact the parameter goodness-of-fit reported in Tab.~\ref{010} shows a tension between
Gallium and Chooz data under the hypothesis of
neutrino oscillations.

Figure~\ref{030} shows
the allowed regions in the
$\sin^{2}2\vartheta$--$\Delta{m}^{2}$ plane obtained with the combined least-squares fit of Gallium and Chooz data.
The values of
$\chi^{2}_{\text{min}}$ and Goodness of Fit
and
the best-fit values of the mixing parameters are given in Tab.~\ref{010}.
It is clear that the combined fit is not good,
since the results of Chooz and the Gallium radioactive source experiments
are in contradiction regarding neutrino disappearance.
The marginal allowed ranges of
$\sin^{2}2\vartheta$ and $\Delta{m}^{2}$ in Tab.~\ref{021}
are of little interest,
since the minima of the corresponding $\Delta\chi^{2}$'s in Fig.~\ref{030}
are very shallow,
except for the upper bound on $\sin^{2}2\vartheta$
driven by Chooz data.
As one can see from the allowed regions in the
$\sin^{2}2\vartheta$--$\Delta{m}^{2}$ plane in Fig.~\ref{030},
the Chooz bound on $\sin^{2}2\vartheta$ in Eq.~(\ref{020})
is weakened by the results of the Gallium radioactive source experiments
in a significant way only for
$\Delta{m}^{2} \gtrsim 10^{-1} \, \text{eV}^{2}$
at the $1\sigma$ level.

\begin{figure}[t!]
\begin{center}
\includegraphics*[bb=23 144 572 704, width=\textwidth]{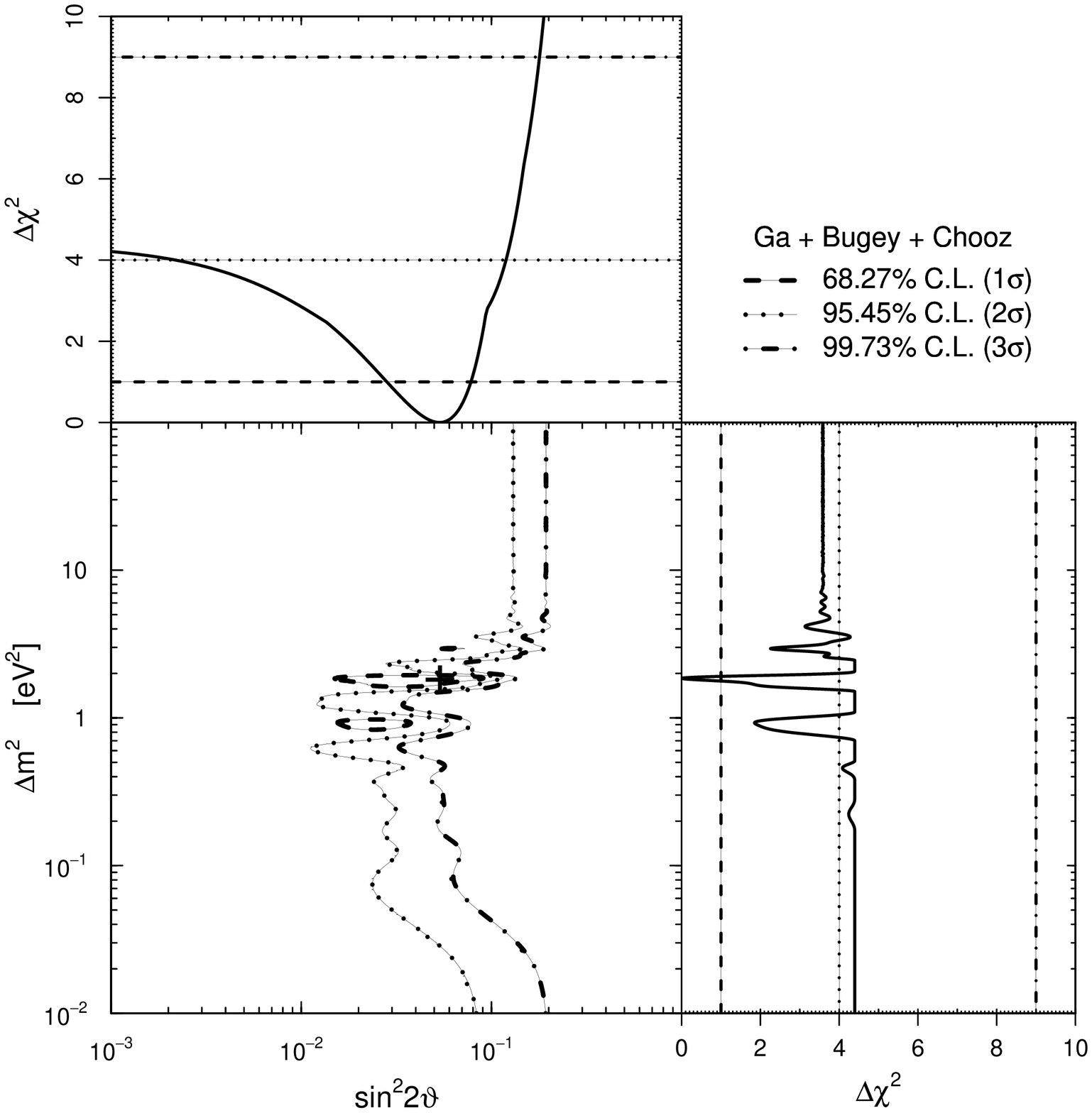}
\caption{ \label{031}
Allowed regions in the
$\sin^{2}2\vartheta$--$\Delta{m}^{2}$ plane
and
marginal $\Delta\chi^{2}$'s
for
$\sin^{2}2\vartheta$ and $\Delta{m}^{2}$
obtained from the
combined fit of the results of
the two GALLEX ${}^{51}\text{Cr}$ radioactive source experiments,
the SAGE
${}^{51}\text{Cr}$ and ${}^{37}\text{Ar}$ radioactive source experiments
and the Bugey and Chooz reactor experiments.
The best-fit point corresponding to $\chi^2_{\text{min}}$ is indicated by a cross.
}
\end{center}
\end{figure}

Finally,
we performed a combined fit of the results of Bugey, Chooz, and Gallium data.
The resulting allowed regions in the
$\sin^{2}2\vartheta$--$\Delta{m}^{2}$ plane
are shown in Fig.~\ref{031}.
The best fit values and the marginal allowed ranges of the mixing parameters
are listed, respectively, in Tabs.~\ref{010} and \ref{021}.
One can see that the Gallium and Chooz data tend to compensate each other,
leading to results which are similar to those obtained in the analysis of Bugey data alone.
The value of the parameter goodness-of-fit reported in Tab.~\ref{010} does not allow us
to exclude the compatibility of the Bugey, Chooz and Gallium data under the hypothesis of neutrino oscillations.
Also the goodness-of-fit obtained in the case of no oscillations,
given in Tab.~\ref{010}, is acceptable.
Therefore,
we can only conclude that
the combined analysis of all the experimental data that we have considered
is compatible both with the case of no oscillations
and with the hint in favor of neutrino oscillations
with
$0.02 \lesssim \sin^{2}2\vartheta \lesssim 0.08$
and
$\Delta{m}^{2} \approx 1.8 \, \text{eV}^{2}$
found in the analysis of Bugey data.

\section{Conclusions}
\label{022}
\nopagebreak

We interpreted the deficit
observed in the Gallium radioactive source experiments
as a possible indication of
the disappearance of electron neutrinos.
We have analyzed the data
in the effective framework of two-neutrino mixing,
which describes neutrino oscillations due to a $ \Delta{m}^{2} $
that is much larger than the solar and atmospheric ones.
We found that
there is an indication of electron neutrino disappearance
due to neutrino oscillations with
$\sin^{2}2\vartheta \gtrsim 0.03$
and
$\Delta{m}^{2} \gtrsim 0.1 \, \text{eV}^{2}$.
We have also studied the compatibility of the data of the Gallium radioactive source experiments
with the data of the Bugey and Chooz reactor short-baseline antineutrino disappearance experiments
in the same effective framework of two-neutrino mixing,
in which the disappearance of neutrinos and antineutrinos are equal.
We found that the Bugey data present a hint of neutrino oscillations
with
$0.02 \lesssim \sin^{2}2\vartheta \lesssim 0.08$
and
$\Delta{m}^{2} \approx 1.8 \, \text{eV}^{2}$,
which is compatible with the region of the mixing parameters
allowed by the analysis of the data of the Gallium radioactive source experiments.
We have also performed combined analyses
of the Bugey and Chooz data,
of the Gallium and Bugey data,
of the Gallium and Chooz data,
which show that 
the Bugey and Chooz data are compatible,
the Gallium and Bugey data are compatible, and
the Gallium and Chooz data are marginally compatible.
The weak indication in favor of neutrino oscillations
found in the analysis of the Bugey data persists in the combined analyses
of the Bugey data with the Gallium and Chooz data.
However,
we cannot exclude the absence of oscillations.

From a physical point of view,
a hint in favor of short-baseline neutrino oscillations
generated by $\Delta{m}^{2} \gtrsim 0.1 \, \text{eV}^{2}$
is extremely interesting.
This squared-mass difference
is too large to be compatible with  the three-neutrino mixing scheme
inferred from the observation of neutrino oscillations in
solar, very-long-baseline reactor, atmospheric and long-baseline accelerator experiments,
in which there are only two independent squared-mass differences,
$ \Delta{m}^{2}_{\text{SOL}} \approx 8 \times 10^{-5} \, \text{eV}^{2} $
and
$ \Delta{m}^{2}_{\text{ATM}} \approx 3 \times 10^{-3} \, \text{eV}^{2} $.
Therefore,
the results of our analysis indicate the possible existence of at least one
light sterile neutrino $\nu_{s}$
(see Refs.~\protect\cite{hep-ph/9812360,hep-ph/0202058,Giunti-Kim-2007}).
We think that it is very important to explore
this intriguing hint of
new physics beyond the Standard Model.

As already discussed in Ref.~\cite{0707.4593},
short-baseline $\boss{\nu}{e}\to\boss{\nu}{s}$ transitions
have an influence on the interpretation of all experiments
with an initial $\boss{\nu}{e}$ beam.
In the existing
solar and atmospheric neutrino experiments the survival probability of $\boss{\nu}{e}$ is
the averaged one in Eq.~(\ref{019}).
However, the uncertainties of the experimental data and our knowledge of the initial flux
do not allow us to exclude $\boss{\nu}{e}\to\boss{\nu}{s}$ transitions at the level of about 20\%
in the case of solar neutrinos
\cite{hep-ph/0406294}
and about 30\%
(see Ref.~\cite{Giunti-Kim-2007})
in the case of atmospheric neutrinos.

Future experiments which are well suited for finding small
$\boss{\nu}{e}\to\boss{\nu}{s}$ transitions
are those with a source producing a $\boss{\nu}{e}$ flux
which is known with high accuracy.
Since sterile neutrinos are invisible,
$\boss{\nu}{e}\to\boss{\nu}{s}$ transitions
can be revealed either by measuring a disappearance of
$\boss{\nu}{e}$'s without $\bos{\mu}$ or $\bos{\tau}$ production in the detection process
or by measuring a disappearance of
$\boss{\nu}{e}$'s due to oscillations with a squared-mass difference
much larger than
$ \Delta{m}^{2}_{\text{SOL}} $
and
$ \Delta{m}^{2}_{\text{ATM}} $.
We are aware of the following possibilities:
Beta-Beam experiments \protect\cite{Zucchelli:2002sa}
which have a pure $\nu_{e}$ or $\bar\nu_{e}$ beam from nuclear decay
(see the reviews in Refs.~\protect\cite{physics/0411123,hep-ph/0605033});
Neutrino Factory experiments
in which the beam is composed of
$\nu_{e}$ and $\bar\nu_{\mu}$,
from $\mu^{+}$ decay,
or
$\bar\nu_{e}$ and $\nu_{\mu}$,
from $\mu^{-}$ decay
(see the review in Ref.~\protect\cite{hep-ph/0210192,physics/0411123});
Mossbauer neutrino experiments,
with a $\bar\nu_{e}$ beam
produced in recoiless nuclear decay
and detected in recoiless nuclear antineutrino capture
\protect\cite{hep-ph/0601079};
the LENS detector
\protect\cite{Raghavan:1997ad,LENS-2002}
with an artificial Megacurie $\nu_{e}$ source
\protect\cite{Grieb:2006mp}.
Let us also notice the very interesting possibility to reveal
the existence of sterile neutrinos in the flux of
high-energy astrophysical neutrinos after their passage through the Earth
by measuring the peculiar matter effects \cite{hep-ph/0302039,0709.1937}.

\section*{Acknowledgments}
\nopagebreak

We would like to express our gratitude to Y. Declais for giving us detailed information
on the Bugey experiment.
M.A. Acero would like to thank the International Doctorate on AstroParticle Physics
(IDAPP) for financial support.
C. Giunti would like to thank the Department of Theoretical Physics of the University of Torino
for hospitality and support.

\raggedright

\end{document}